\documentclass[aps,prd,reprint,floatfix,superscriptaddress,amsmath,amssymb]{revtex4-2}
\usepackage[T1]{fontenc}
\usepackage[utf8]{inputenc}
\usepackage{graphicx}
\usepackage{bm}
\usepackage[hidelinks]{hyperref}

\begin{document}
\raggedbottom

\title{Next-to-leading-order QCD corrections to prompt $\Upsilon(1S)$ pair hadroproduction}
\author{Zhi-Guo He}
\email{zhiguo.he@buct.edu.cn}
\affiliation{College of Mathematics and Physics, Beijing University of Chemical Technology, Beijing 100029, China}

\author{Xiao-Bo Jin}
\email{xiaobojinpro@gmail.com}
\affiliation{College of Mathematics and Physics, Beijing University of Chemical Technology, Beijing 100029, China}
\author{Bernd A. Kniehl}
\email{kniehl@mail.desy.de}
\affiliation{{II.} Institut f\"ur Theoretische Physik, Universit\"at Hamburg, Luruper Chaussee 149, 22761 Hamburg, Germany}

\begin{abstract}
  Working at next-to-leading order in nonrelativistic-QCD factorization, we study prompt $\Upsilon(1S)$ pair hadroproduction, including feed-down from the excited $\Upsilon(2S)$ and $\Upsilon(3S)$ states and focusing on the color singlet contributions, which are known to be dominant. 
Besides the integrated cross section $\sigma^{\Upsilon\Upsilon}$, we predict distributions in the invariant mass $m_{\Upsilon\Upsilon}$, transverse momentum $p_T^{\Upsilon\Upsilon}$, and rapidity separation $|\Delta y|$ of the $\Upsilon(1S)$ pair, as well as the one in the transverse momentum $p_T^{\Upsilon}$ of either $\Upsilon(1S)$ meson, and compare them with measurements by the CMS Collaboration at the CERN LHC with center-of-mass energies $\sqrt{s}=8$ and $13~\mathrm{TeV}$ as far as they go.
The QCD corrections to $\sigma^{\Upsilon\Upsilon}$ turn out to be mildly negative, with $K$ factors of about 0.9.
By contrast, the differential cross sections receive $K$ factors as small as 0.6 and 0.2 at large values of $m_{\Upsilon\Upsilon}\le 100$~GeV and $|\Delta y|\le4$, respectively, and as large as 20 for large values of $p_T^{\Upsilon}\le30$~GeV.
Our predictions agree well with the data, except in the outermost bins of $m_{\Upsilon\Upsilon}$ and $|\Delta y|$, where they fall short of the data, leaving room for contributions from double parton scattering as expected.
Our analysis thus provides a first solid basis for the extraction of the key parameter $\sigma_{\mathrm{eff}}$ of double parton scattering.
\end{abstract}
\maketitle

\section{Introduction}
\label{sec:introduction}

Bottomonia, the bound states of a bottom quark and its antiquark, occupy a special place in heavy-quarkonium physics~\cite{Brambilla:2010cs}.
Since the typical relative velocity of the constituent quarks is smaller inside bottomonium than inside charmonium, the velocity expansion for bottomonium production is expected to converge better within the framework of nonrelativistic-QCD (NRQCD) factorization~\cite{Bodwin:1994jh}.
Heavy-quarkonium pair production additionally involves the transitions of two heavy quark-antiquark pairs into physical quarkonium states.
These features make double $\Upsilon$ hadroproduction a particularly valuable probe of the nonperturbative dynamics of quarkonium formation~\cite{Qiao:2002rh,Li:2009ug,Ko:2010xy}.
In addition to the conventional single parton scattering (SPS) mechanism, the $\Upsilon(1S)$ pair can also be produced through double parton scattering (DPS), in which the two quarkonia originate from separate hard partonic interactions in the same proton-proton collision.
Under the usual factorized treatment of DPS, the pair production cross section can be expressed as 
\begin{equation}
\sigma_{\mathrm{DPS}}^{\Upsilon\Upsilon}=\frac{\sigma^{\Upsilon}\sigma^{\Upsilon}}{2\sigma_{\mathrm{eff}}}\,,
\end{equation} 
where the parameter $\sigma_{\mathrm{eff}}$ is related to the transverse distribution of partons inside the proton~\cite{Kom:2011bd}.

The CMS Collaboration first observed $\Upsilon(1S)$ pair production in proton-proton collisions at center-of-mass energy $\sqrt{s}=8~\mathrm{TeV}$ in the fiducial region $|y^{\Upsilon(1S)}|<2.0$ of rapidity for each meson~\cite{CMS:2016liw}.
In 2020, using a much larger data set collected at $\sqrt{s}=13~\mathrm{TeV}$ in the same fiducial region, CMS reported not only the integrated cross section $\sigma^{\Upsilon\Upsilon}$, but also differential cross sections with respect to the invariant mass $m_{\Upsilon\Upsilon}$ and rapidity difference $|\Delta y|$~\cite{CMS:2020qwa}.
Moreover, CMS attempted to separate the SPS and DPS contributions through $m_{\Upsilon\Upsilon}$ and $|\Delta y|$ distributions, and searched for resonances decaying to $\Upsilon(1S)\mu^+\mu^-$ as possible fully-bottom tetraquark candidates~\cite{CMS:2020qwa}.

On the theoretical side, pioneering leading-order (LO) studies of $\Upsilon(1S)$ pair hadroproduction within NRQCD considered the color-singlet (CS) channel and one specific color-octet (CO) contribution \cite{Li:2009ug,Ko:2010xy}.
Higher-order relativistic corrections at relative order $\mathcal{O}(v^2)$ to these channels were subsequently investigated~\cite{Li:2013csa}.
$\Upsilon(1S)$-pair production has also been studied in the color-evaporation model \cite{Lansberg:2020rft}.
However, a full next-to-leading-order (NLO) NRQCD analysis of double $\Upsilon(1S)$ hadroproduction in SPS has so far been lacking.

A more detailed picture of quarkonium pair production has emerged from studies of double $J/\psi$ hadroproduction.
Early work proposed gluon fragmentation through CO states as a probe of the CO mechanism at large transverse momentum $p_T^{\psi}$~\cite{Barger:1995vx}, while calculations of the CS channel established its importance at low and moderate values of $p_T^{\psi}$~\cite{Qiao:2002rh,Li:2009ug,Ko:2010xy}.
A complete LO survey of prompt double $J/\psi$ hadroproduction subsequently showed that interactions of the $t$-channel gluon-exchange type can enhance CO and $\chi_{cJ}$ feed-down contributions compared to the double CS channel, making them predominant at large $m^{\psi\psi}$ and $|\Delta y^{\psi\psi}|$ within the CMS and ATLAS acceptance ranges~\cite{He:2015qya}.
Our recent $\mathcal{O}(\alpha_s^5)$ analysis in the large-$p_T^{\psi\psi}$ region has yielded complementary constraints on CO long-distance matrix elements (LDMEs)~\cite{He:2026odn}.

Several studies addressed higher-order corrections in both $v^2$ and $\alpha_s$.
Relativistic effects at subleading order in $v^2$ were investigated in Refs.~\cite{Li:2013csa,He:2024ugx}.
A dedicated analysis near the threshold region found that retaining the physical phase space together with these corrections improved the description of the LHCb measurements of the invariant mass distribution~\cite{He:2024ugx}.
The effects of higher-order QCD radiative corrections were first studied in the NLO$^{\star}$ approximation for the CS channel, in which only real hard-gluon emission was included, with a cutoff used to regularize the infrared divergences~\cite{Lansberg:2014swa,Lansberg:2013qka}.
The first complete NLO calculation of the direct CS contribution was presented in Ref.~\cite{Sun:2014gca}, and the ${}^1S_0^{[8]}+{}^1S_0^{[8]}$ channel was subsequently studied at NLO in Ref.~\cite{Sun:2023exa}.
Our independent NLO CS calculation~\cite{He:2025kkw} has enabled a systematic comparison with the LHCb, CMS, and ATLAS measurements.
We found that the NLO-QCD CS contribution remains below the double $J/\psi$ cross sections measured by CMS and ATLAS, imposing sizable lower cuts on $p_T^{J/\psi}$.
However, under the LHCb cuts, which only imply an upper bound on $p_T^{J/\psi}$ and thus retain the low $p_T^{J/\psi}$ region, the CS contribution can account for the data in most bins.

Unlike the double $J/\psi$ measurements, the CMS measurements of $\Upsilon(1S)$ pair production imposed no lower cut on $p_T^{\Upsilon}$~\cite{CMS:2016liw,CMS:2020qwa}.
With both mesons restricted to $|y^{\Upsilon}|<2.0$, the CMS integrated cross sections mainly sample the small-$p_T^{\Upsilon}$ region, in which the double CS channel is expected to dominate the direct SPS yield \cite{Ko:2010xy}.
The prominence of this region connects the CMS bottomonium measurements to the LHCb double $J/\psi$ results, for which the NLO CS predictions describe the data in most bins rather well~\cite{He:2025kkw}.
This observation motivates a quantitative study of the CS SPS contribution and its kinematic dependences in the CMS double-$\Upsilon$ measurements.
An improved CS SPS prediction would also facilitate the extraction of the DPS fraction in the kinematic fits~\cite{CMS:2020qwa}.

In this work, we present what we believe to be the first full NLO-QCD calculation of the CS SPS contribution to $\Upsilon(1S)$ pair hadroproduction within NRQCD factorization.
Our phenomenological predictions also include the feed-down from the $\Upsilon(2S)$ and $\Upsilon(3S)$ mesons.
We evaluate $\sigma^{\Upsilon\Upsilon}$ at $\sqrt{s}=8$ and $13~\mathrm{TeV}$ as well as four differential distributions, in $m_{\Upsilon\Upsilon}$, $|\Delta y|$, $p_T^{\Upsilon}$, and $p_T^{\Upsilon\Upsilon}$, at $13~\mathrm{TeV}$ adopting the respective CMS kinematic conditions \cite{CMS:2016liw,CMS:2020qwa}.
We compare our predictions with the CMS measurements of $\sigma^{\Upsilon\Upsilon}$~\cite{CMS:2016liw,CMS:2020qwa} and the $m_{\Upsilon\Upsilon}$ and $|\Delta y|$ distributions \cite{CMS:2020qwa}.
This analysis quantifies the effects of higher-order QCD radiation on the normalization and shapes of the CS SPS contribution.
It provides input for future studies of DPS once the remaining SPS contributions are assessed, and may also help to model the nonresonant $\Upsilon(1S)$ pair background in fully-bottom tetraquark searches.

The remainder of this paper is organized as follows.
Section~\ref{sec:theory} describes the theoretical framework for the NLO QCD calculation, the treatment of the virtual and real corrections, and the test of slicing-parameter stability.
Section~\ref{sec:phenomenology} presents the numerical results for the integrated cross sections and four differential distributions and discusses their phenomenological implications.
Section~\ref{sec:conclusions} contains our conclusions.

\section{Theoretical framework}

\label{sec:theory}
In this section, we outline the theoretical framework utilized to compute the cross section for 
prompt $\Upsilon(1S)$ pair hadroproduction. Invoking the collinear parton model and NRQCD 
factorization, the differential cross section for prompt $\Upsilon(1S)$ pair hadroproduction 
through the SPS mechanism, including feed-down contributions, can be written as 
\begin{widetext}
\begin{eqnarray}\label{eq:factorization}
&\mathrm{d}\sigma_{\mathrm{SPS}}^{\mathrm{prompt}}
\bigl(pp\to 2\Upsilon(1S)+X\bigr)
={}\sum_{i,j,m,n}\sum_{H_1,H_2}
\int_0^1\mathrm{d}x_1\,\mathrm{d}x_2\,
f_{i/p}(x_1,\mu_f)f_{j/p}(x_2,\mu_f)\nonumber\\
&\times\mathrm{d}\hat{\sigma}\bigl(i+j\to b\bar b(m)+b\bar b(n)+X\bigr)
\langle\mathcal{O}^{H_1}(m)\rangle
\langle\mathcal{O}^{H_2}(n)\rangle\times\operatorname{Br}\bigl(H_1\to\Upsilon(1S)+X\bigr)
\operatorname{Br}\bigl(H_2\to\Upsilon(1S)+X\bigr)\,.
\end{eqnarray}
\end{widetext}
Here $f_{i/p}$ denotes the parton distribution function (PDF) for parton $i$ within the proton, 
$x_{1,2}$ are the corresponding longitudinal momentum fractions, $\mu_f$ is the factorization 
scale, and $\mathrm{d}\hat\sigma$ is the parton-level short-distance coefficient evaluated 
using the covariant projection method~\cite{Bodwin:2002cfe}. The NRQCD LDMEs 
$\langle\mathcal{O}^{H}(n)\rangle$ encode the nonperturbative transition of the $b\bar{b}$ pair 
from a specific Fock state $n$ into the physical meson $H$, where $n$ specifies the 
spectroscopic notation ${}^{2S+1}L_J^{[c]}$, with $c=1,8$ for CS and CO, respectively. The 
inclusive branching fraction $\operatorname{Br}\bigl(H\to\Upsilon(1S)+X\bigr)$ weights the feed-down contribution from each parent meson and is set to unity for $H=\Upsilon(1S)$. In this work, we focus only on the CS channel with $m=n={}^3S_1^{[1]}$ and the parent states 
$H_{1,2}=\Upsilon(1S),\Upsilon(2S),\Upsilon(3S)$. 

At LO, we retain the gluon-fusion subprocess
$g+g\to b\bar b({}^3S_1^{[1]})+b\bar b({}^3S_1^{[1]})$,
of order $\alpha_s^4$, and neglect quark-antiquark annihilation because of its small parton luminosity. The derivation of the analytic results for NLO calculation has been described in detail in Ref.~\cite{He:2025kkw} for the analogous case of $2c\bar c({}^3S_1^{[1]})$. They can be used here, too, after appropriately adjusting the inputs. The only difference between Ref.~\cite{He:2025kkw} and the case at hand is that we are dealing here with two massive quark flavors, charm and bottom. This is reflected in the renormalization constants. The corresponding counterterms we used here are
\begin{widetext}
\begin{subequations}\label{eq:renormalization}
\begin{align}
\delta Z_2^{\mathrm{OS}}
&=-C_F\frac{\alpha_s}{4\pi}
N_\epsilon
\left(
  \frac{1}{\epsilon_{\mathrm{UV}}}
  +\frac{2}{\epsilon_{\mathrm{IR}}}
  +4
\right)+\mathcal{O}(\alpha_s^2)\,,\\
\delta Z_m^{\mathrm{OS}}
&=
-3C_F\frac{\alpha_s}{4\pi}
N_\epsilon
\left(
  \frac{1}{\epsilon_{\mathrm{UV}}}
  +\frac{4}{3}
\right)+\mathcal{O}(\alpha_s^2)\,,\\
\delta Z_3^{\mathrm{OS}}
&=
\frac{\alpha_s}{4\pi}
N_\epsilon
\left[(\beta_0^\prime-2C_A)
\left(
  \frac{1}{\epsilon_{\mathrm{UV}}}
  -\frac{1}{\epsilon_{\mathrm{IR}}}
\right)-\frac{4}{3}
T_F
\left(
\frac{(n_f-n_l)}{\epsilon_{\mathrm{UV}}}
+
\ln \frac{m_b^2}{m_c^2}
\right)
\right]+\mathcal{O}(\alpha_s^2)\,,\\
\delta Z_g^{\overline{\mathrm{MS}}}
&=
-\frac{\beta_0}{2}\,
\frac{\alpha_s}{4\pi}
N_\epsilon
\left(
  \frac{1}{\epsilon_{\mathrm{UV}}}
  +\ln\frac{m_b^2}{\mu_r^2}
\right)+\mathcal{O}(\alpha_s^2)\,,
\end{align}
\end{subequations}
\end{widetext}
where $N_\epsilon=(4\pi\mu_r^2/m_b^2)^\epsilon/\Gamma(1-\epsilon)$, 
and $\alpha_s\equiv\alpha_s(\mu_r)$ with $\mu_r$ denoting the renormalization scale; the color factors are $C_A=3$, $C_F=4/3$, and $T_F=1/2$, and the beta function coefficients are 
$\beta_0=(11C_A-4T_Fn_f)/3$ and $\beta_0^\prime=(11C_A-4T_Fn_l)/3$. We take $n_f=5$ and $n_l=3$, so that the massive quark contribution to 
$\delta Z_3^{\mathrm{OS}}$ comes from both the charm and bottom quarks. 

The $2\to3$ real corrections of the $gg$ subprocess contain both soft and hard 
collinear singularities in the phase space integration, whereas the $gq(\bar{q})$ 
subprocesses only include hard collinear singularities. Here we isolate these soft and hard collinear contributions by partitioning the phase space using the two-cutoff phase-space-slicing method~\cite{Harris:2001sx} in the same way as that in Ref.~\cite{He:2025kkw}.

In the two-cutoff phase-space-slicing method, the cross section of the 
real corrections is divided into three pieces -- the soft part ($\mathrm{S}$), the hard collinear part $(\mathrm{HC})$, and the hard noncollinear part 
($\overline{\mathrm{HC}}$), as
\begin{equation}
   \sigma^{\mathrm{real}}= \sigma^{\mathrm{S}}+\sigma^{\mathrm{HC}}+\sigma^{\overline{\mathrm{HC}}}\,.
\end{equation}
Each of the three terms is finite but depends on the parameters $\delta_s$ and $\delta_c$, 
while their sum, $\sigma^{\mathrm{real}}$, is independent of them. Since the cancellation of the $\delta_s$ and $\delta_c$ dependence is performed numerically, 
their values must be chosen carefully to ensure the convergence and stability 
of the numerical results. Note that we can not taken those values used in double 
$J/\psi$ hadroproduction directly.

\begin{figure*}[tbp]
\centering
\begin{minipage}[t]{0.48\textwidth}
\centering
\includegraphics[width=\linewidth]{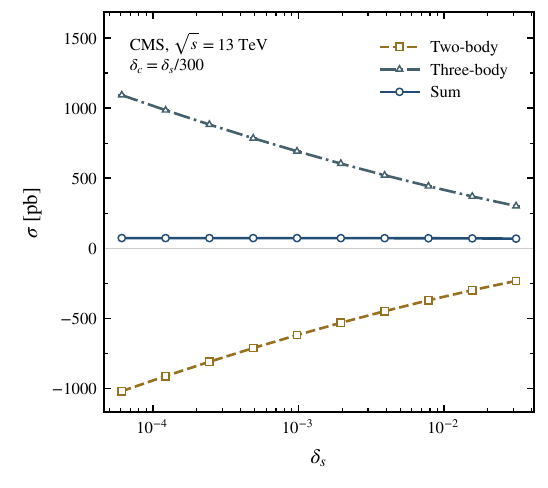}\\[-0.5ex]
(a)
\end{minipage}\hfill
\begin{minipage}[t]{0.48\textwidth}
\centering
\includegraphics[width=\linewidth]{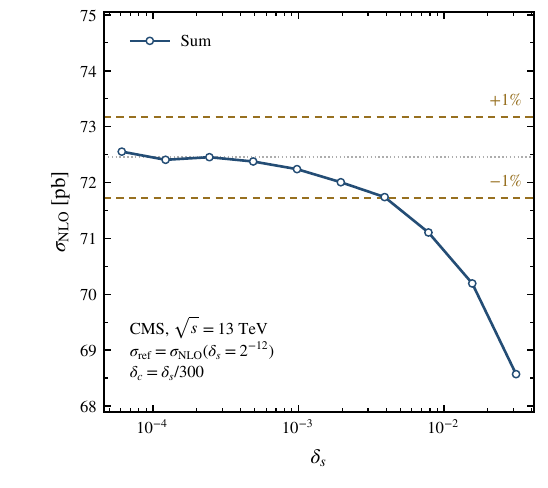}\\[-0.5ex]
(b)
\end{minipage}
\caption{Cutoff dependence of the integrated cross section for the CS contribution in SPS for 
$\Upsilon(1S)$ pair production at $\sqrt{s}=13~\mathrm{TeV}$, with $\delta_c=\delta_s/300$. 
Panel (a) displays the sum alongside its two-body and three-body components. Panel (b) 
magnifies the residual variation of the sum with $\delta_s$. The result shown by the dotted 
lines is computed at $\delta_s=2^{-12}$, and the dashed lines indicate $\pm1\%$ deviations from 
this reference.}
\label{fig:cutoff}
\end{figure*}

We choose the integrated cross section for the CMS 13~TeV measurement as a test by varying $\delta_s$ from $2^{-14}$ to $2^{-5}$ and fixing $\delta_c=\delta_s/300$. For 
illustration, Fig.~\ref{fig:cutoff}(a) displays the contributions grouped according to two-body 
and three-body kinematics, where the two-body contribution includes virtual 
corrections, $\sigma^{\mathrm{S}}$ and $\sigma^{\mathrm{HC}}$, while the three-body contribution contains only $\sigma^{\overline{\mathrm{HC}}}$. As shown in Fig.~\ref{fig:cutoff}(a), their 
strong cutoff dependencies largely cancel in the complete sum. Figure~\ref{fig:cutoff}(b) 
shows more clearly that the NLO cross section approaches a stable plateau as 
$\delta_s$ decreases. We observe that the integrated cross section becomes sufficiently stable below 
$\delta_s=2^{-10}$ with $\delta_c=\delta_s/300$. We therefore fix their values at 
$\delta_s=2^{-12}$ and $\delta_c=\delta_s/300$ throughout the numerical calculations.  

\section{Phenomenology}
\label{sec:phenomenology}

We are now in a position to explore the phenomenological implications of our study.
We start by specifying the parameter inputs for the numerical evaluations.
At LO (NLO), we adopt set CTEQ6L1 (CTEQ6M) of the proton's parton distribution functions~\cite{Pumplin:2002vw} in combination with one-loop (two-loop) evolution of the QCD coupling constant $\alpha_s^{n_f}(\mu_r)$, with $n_f=5$ active quark flavors and asymptotic scale parameter $\Lambda_{\mathrm{QCD}}^{(5)}=165~\mathrm{MeV}$ ($226~\mathrm{MeV}$).
The heavy-quark masses are set to $m_c=1.5~\mathrm{GeV}$ and $m_b=4.75~\mathrm{GeV}$. 
We choose the renormalization and factorization scales to be $\mu_r=\mu_f=\xi\mu_0$, where $\mu_0=(m_{T_1}+m_{T_2})/2$ and $m_{T_i}=\sqrt{4m_b^2+(p_T^{\Upsilon_i})^2}$, with $i=1,2$ numbering the $\Upsilon(1S)$ mesons.
Since the mass differences among the $\Upsilon(1S)$, $\Upsilon(2S)$, and $\Upsilon(3S)$ mesons are less than ten percent, we neglect parent-state mass differences in the kinematics.
Thus the SDCs are independent of the parent-state labels.
The theoretical uncertainties are estimated by varying $\xi$ from $1/2$ to $2$ about the default value $\xi=1$.
For $S$-wave states, the NRQCD CS LDMEs can be related to the wave function at the origin through 
\begin{equation}
    \langle\mathcal{O}^{\Upsilon(nS)}({}^3S_1^{[1]})\rangle=\frac{9|R_{nS}(0)|^2}{2\pi}\,.
\end{equation}
We adopt the values evaluated in Ref.~\cite{Eichten:1995ch} using the Buchm\"uller--Tye potential, yielding
\begin{eqnarray}\label{eq:ldmes}
  \langle\mathcal{O}^{\Upsilon(1S)}({}^3S_1^{[1]})\rangle&=&9.278~\mathrm{GeV}^3\,,
  \nonumber\\
  \langle\mathcal{O}^{\Upsilon(2S)}({}^3S_1^{[1]})\rangle&=&4.632~\mathrm{GeV}^3\,,
  \nonumber\\
\langle\mathcal{O}^{\Upsilon(3S)}({}^3S_1^{[1]})\rangle&=&3.544~\mathrm{GeV}^3\,.
\end{eqnarray}
We take the $\Upsilon(2S)$ and $\Upsilon(3S)$ to $\Upsilon(1S)$ decay branching fractions from the Particle Data Group~\cite{ParticleDataGroup:2026mpi}.
Additionally, for $\Upsilon(3S)$, the cascade decay $\Upsilon(3S)\to\Upsilon(2S)\to\Upsilon(1S)$ is taken into account.
Summing over the predominant channels of hadronic transitions and double $E1$ transitions, we get 
\begin{eqnarray}
  Br(\Upsilon(2S)\to\Upsilon(1S)+X)&=& 30.3\%\,,
\nonumber\\
Br(\Upsilon(3S)\to\Upsilon(1S)+X)&=& 12.6\%\,.
\end{eqnarray}

\begin{table*}[tbp]
  \caption{The integrated cross sections of double prompt $\Upsilon(1S)$ hadroproduction measured by CMS at $\sqrt{s}=8$ and 13~TeV in the $|y^{\Upsilon}|<2.0$ fiducial region for each meson \cite{CMS:2016liw,CMS:2020qwa} are compared with our predictions evaluated at LO and NLO in $\alpha_s$.
The central values refer to $\xi=1$; the scale-variation bounds are set by the predictions for $\xi=1/2$ and $\xi=2$.
    The ratio
$K\equiv\sigma^{\mathrm{NLO}}_{\mathrm{CS}}/\sigma^{\mathrm{LO}}_{\mathrm{CS}}$
    is evaluated at the same value of $\xi$ at LO and NLO.
    The CMS uncertainties are statistical, systematic, and due to the dimuon branching fraction $\mathcal{B}=\operatorname{Br}(\Upsilon(1S)\to\mu^+\mu^-)$.}
\label{tab:cross-sections}
\begin{ruledtabular}
\begin{tabular}{ccccc}
$\sqrt{s}$ & CMS [pb] & LO [pb] & NLO [pb] & $K$ \\
\hline
$8~\mathrm{TeV}$ & $68.8\pm12.7(\mathrm{stat})\pm7.4(\mathrm{syst})\pm2.8(\mathcal{B})$ & $49.3^{+19.9}_{-14.8}$ & $45.5^{+24.8}_{-12.2}$ & $0.922^{+0.093}_{+0.043}$ \\
$13~\mathrm{TeV}$ & $79\pm11(\mathrm{stat})\pm6(\mathrm{syst})\pm3(\mathcal{B})$ & $80.2^{+25.5}_{-21.3}$ & $72.5^{+36.8}_{-18.6}$ & $0.903^{+0.131}_{+0.012}$ \\
\end{tabular}
\end{ruledtabular}
\end{table*}

We start by considering the integrated cross section of double prompt $\Upsilon(1S)$ hadroproduction in SPS.
Both CMS measurements refer to the kinematic region $|y^{\Upsilon}|<2.0$ for each meson, without imposing a $p_T^{\Upsilon}$ cut on either $\Upsilon(1S)$, and assume unpolarized $\Upsilon(1S)$ mesons~\cite{CMS:2016liw,CMS:2020qwa}. 
The CS predictions, including feed-down, at LO and NLO in $\alpha_s$ are summarized in Table~\ref{tab:cross-sections} along with the CMS measurements at $\sqrt{s}=8~\mathrm{TeV}$ and $\sqrt{s}=13~\mathrm{TeV}$~\cite{CMS:2016liw,CMS:2020qwa}.
At both center-of-mass energies, the CS contributions are compatible with the CMS measurements within the theoretical and experimental uncertainties.
In the $8~\mathrm{TeV}$ case, the central LO prediction is $49.3~\mathrm{pb}$, and the NLO QCD corrections reduce it by about $7.8\%$ to $45.5~\mathrm{pb}$, which is about $66.1\%$ of the experimental central value of $68.8~\mathrm{pb}$, but remains compatible once the scale variations are included.
A similar behavior is observed in the $13~\mathrm{TeV}$ case.
The central LO prediction of $80.2~\mathrm{pb}$ is shifted to $72.5~\mathrm{pb}$ at NLO, and it is in good agreement with the measured value of $79 \pm 13~\mathrm{pb}$.
The $K$ factors, $K=\sigma_{\mathrm{CS}}^{\mathrm{NLO}}/\sigma_{\mathrm{CS}}^{\mathrm{LO}}$, obtained at the default scale $(\xi=1)$ lie slightly below unity for both cases, being $K \approx 0.922$ at $\sqrt{s}=8~\mathrm{TeV}$ and $K \approx 0.903$ at $\sqrt{s}=13~\mathrm{TeV}$, indicating that the effect of NLO corrections is small and negative for the integrated cross section.
The scale variations shift $K$ upward by $+0.093/+0.043$ and $+0.131/+0.012$ for the 8 and $13~\mathrm{TeV}$ cases, respectively. 
At both energies, the NLO scale bands overlap with the experimental uncertainty intervals.
However, a reduction in the scale dependence of the integrated cross sections at NLO is not readily apparent.
In contrast, the energy ratio,
$R_{13/8}\equiv\sigma_{\mathrm{CS}}(13~\mathrm{TeV})/\sigma_{\mathrm{CS}}(8~\mathrm{TeV})$,
takes the values $(1.526,\,1.626,\,1.706)$ at LO and $(1.555,\,1.593,\,1.619)$ at NLO for $\xi=1/2,1,2$, respectively.
This demonstrates that the energy ratio is significantly less sensitive to the scale choice at NLO, highlighting a notable improvement in perturbative stability.

\begin{figure*}[tbp]
\begin{minipage}[t]{0.48\textwidth}
\centering
\includegraphics[width=\linewidth]{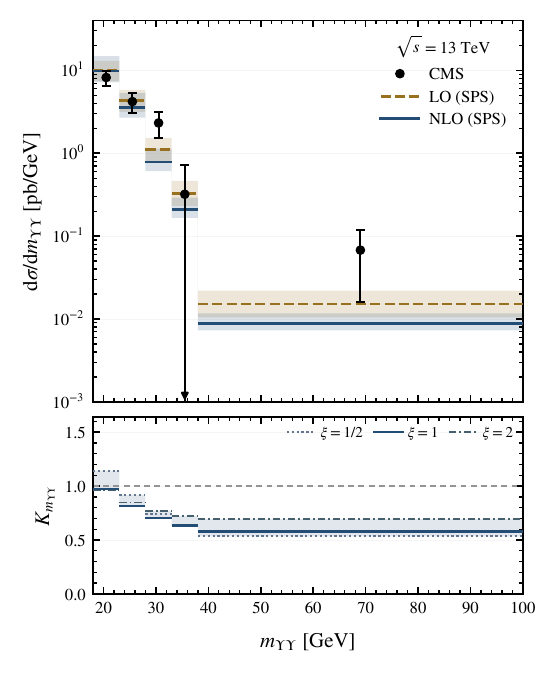}
\caption{The central LO (golden dashed line) and NLO (blue solid line) CS predictions of the $m_{\Upsilon\Upsilon}$ distribution of double prompt $\Upsilon(1S)$ hadroproduction in SPS are compared with the CMS measurements (black points) at $\sqrt{s}=13~\mathrm{TeV}$ \cite{CMS:2020qwa}.
The arrow marks an experimental error bar extending below zero.
The golden and blue bands show the scale uncertainties at LO and NLO, respectively.
The lower panel displays the corresponding differential $K$ factors.}
\label{fig:mpair}
\end{minipage}\hfill
\begin{minipage}[t]{0.48\textwidth}
\centering
\includegraphics[width=\linewidth]{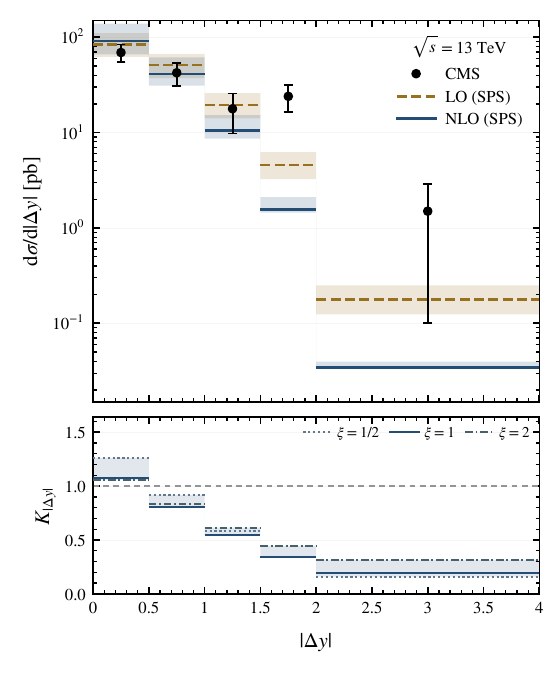}
\caption{Same as in Fig.~\ref{fig:mpair}, but for the $|\Delta y|$ distribution.}
\label{fig:dely}
\end{minipage}
\end{figure*}

\begin{figure*}[tbp]
\begin{minipage}[t]{0.48\textwidth}
\centering
\includegraphics[width=\linewidth]{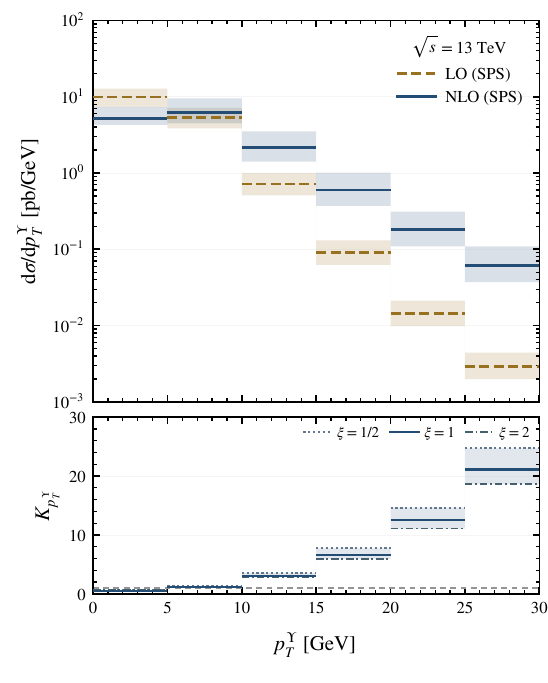}
\caption{Central LO (golden dashed line) and NLO (blue solid line) CS predictions of the $p_T^{\Upsilon}$ distribution of double prompt $\Upsilon(1S)$ hadroproduction in SPS.
The golden and blue bands show the scale uncertainties at LO and NLO, respectively.
The lower panel displays the corresponding differential $K$ factors.}
\label{fig:pt}
\end{minipage}\hfill
\begin{minipage}[t]{0.48\textwidth}
\centering
\includegraphics[width=\linewidth]{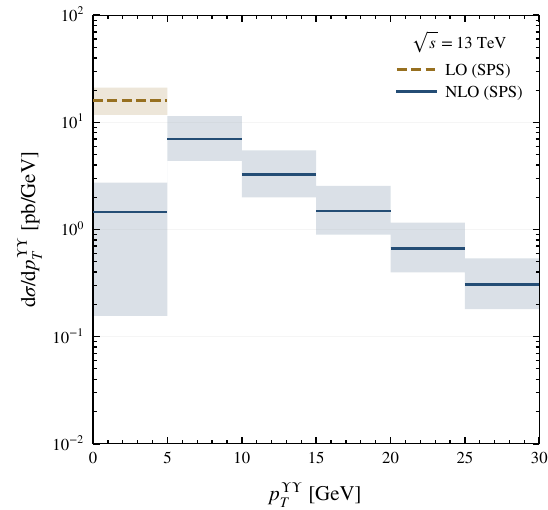}
\caption{
Same as in Fig.~\ref{fig:pt}, but without the $K$ factors.}
\label{fig:ptpair}
\end{minipage}
\end{figure*}

We next examine differential distributions of double prompt $\Upsilon(1S)$ hadroproduction in SPS at $\sqrt{s}=13~\mathrm{TeV}$.
In Figs.~\ref{fig:mpair}, \ref{fig:dely}, and \ref{fig:pt}, we consider the distributions in meson pair invariant mass $m_{\Upsilon\Upsilon}$ and rapidity separation $|\Delta y|$, and in single-meson transverse momentum $p_T^{\Upsilon}$, respectively.
In each case, the upper panels show the CS predictions at LO and NLO in $\alpha_s$ and the lower panels show the $K$ factors evaluated at $\xi=1/2,1,2$, quantifying the effect of the NLO QCD corrections in each bin.
In Fig.~\ref{fig:ptpair}, we study the distribution in meson pair transverse momentum $p_T^{\Upsilon\Upsilon}$, in which case there are no $K$ factors to evaluate because $p_T^{\Upsilon\Upsilon}=0$ at LO.
In Figs.~\ref{fig:mpair} and \ref{fig:dely}, the respective CMS data \cite{CMS:2020qwa} are included for comparison.

From Fig.~\ref{fig:mpair}, we observe that the cross section rapidly falls off as $m_{\Upsilon\Upsilon}$ increases.
The LO predictions agree with the CMS measurements \cite{CMS:2020qwa} in all bins within errors, in the sense that the theoretical and experimental error bands overlap.
For $\xi=1$, the $K$ factor is about $0.98$ in the first bin and decreases down to $0.58$ in the last bin.
In other words, the NLO corrections have little effect on the central prediction in the first bin, but the $m_{\Upsilon\Upsilon}$ distribution falls off more rapidly at NLO than at LO.
The fall-off is more (less) pronounced at $\xi=1/2$ (2).
As a consequence, the NLO uncertainty bands overlap with the CMS error intervals in the first and second bins, but tend to undershoot them in the other bins.
The agreement in the fourth bin has to be taken with a grain of salt in view of the large experimental uncertainty there.
The CMS central value in the highest $m_{\Upsilon\Upsilon}$ bin is approximately $7.7$ times larger than the central NLO prediction.

From Fig.~\ref{fig:dely}, we observe that the LO predictions well describe the CMS measurements \cite{CMS:2020qwa} in all $|\Delta y|$ bins, except in the fourth one, where the experimental data point appears to be an outlier.
The QCD corrections are modest in the first two bins, but significantly reduce the LO predictions in the last two bins, causing the NLO predictions to greatly underestimate the CMS measurements.
For $\xi=1$, the $K$ factor drops sharply from approximately $1.08$ in the first bin to a mere $0.19$ in the last bin.
Consequently, the NLO predictions become more concentrated at small $|\Delta y|$ values than their LO counterparts.
The NLO uncertainty bands overlap with the CMS errors in the first three bins, while they fall significantly short of them in the tail.
In the last bin, the central prediction sits about a factor of $44$ below the CMS central value, the experimental uncertainty being large though.
This is a welcome feature because substantial contributions from the CO channels and DPS are expected in this bin.

From Fig.~\ref{fig:pt}, we observe that the LO predictions rapidly fall off as $p_T$ increases.
The NLO corrections completely change the game, entailing a considerable suppression in the first bin and a substantial enhancement beyond the second bin.
For $\xi=1$, the differential $K$ factor rises sharply from about $0.52$ in the first bin to roughly $21$ in the last bin. 
This trend carries over to $\xi=1/2,2$.
As a consequence, the NLO predictions exhibit a very different lineshape, with a maximum at the second bin and a much harder tail.
Due to the strong $p_T$ dependence of the $K$ factor, the NLO distributions cannot usefully be approximated by rescaling their LO counterparts with an integrated  $K$ factor.

Finally, we turn to the $p_T^{\Upsilon\Upsilon}$ distribution in Fig.~\ref{fig:ptpair}.
For the kinematic reason explained above, the LO cross section is confined to the lowest bin.
By contrast, at NLO, recoil against an additional parton allows for the meson pair to acquire nonzero $p_T$ values, the QCD corrections being entirely due to real emission in the remaining bins.
The large negative QCD correction in the first bin indicates that the fixed-order prediction is unreliable at small $p_T^{\Upsilon\Upsilon}$ values, where soft and collinear radiation generates large logarithms.
A reliable description of this region therefore requires $p_T^{\Upsilon\Upsilon}$ resummation matched to the fixed-order result, which is beyond the scope of our present study.

\section{Conclusions}
\label{sec:conclusions}

We calculated the CS contribution to double $\Upsilon(1S)$ hadroproduction in SPS at full NLO accuracy within NRQCD factorization, including the feed-down from the $\Upsilon(2S)$ and $\Upsilon(3S)$ states.
We compared our predictions with the CMS measurements at $\sqrt{s}=8~\mathrm{TeV}$ \cite{CMS:2016liw} and $13~\mathrm{TeV}$~\cite{CMS:2020qwa}.
As for the integrated cross sections, the effect of the QCD corrections was found to be modest, with central $K$ factors of about $0.9$.
Their inclusion slightly impairs the degree of agreement with the CMS data, but the theoretical and experimental uncertainties still nicely overlap.
However, the NLO corrections produce pronounced changes in differential distributions.
In fact, they considerably suppress the distributions at large $m_{\Upsilon\Upsilon}$ or $|\Delta y|$ values, driving the NLO predictions to lie entirely below the CMS uncertainty bands in several bins.
The largest relative deficit occurs in the last $|\Delta y|$ bin, where the CMS central value is about $44$ times as large as the central NLO prediction, albeit the experimental uncertainty is large there.

Similar discrepancies in double $J/\psi$ hadroproduction have motivated studies of CO channels with $t$-channel gluon exchange~\cite{He:2015qya} and high-energy resummation within the Parton Reggeization Approach~\cite{He:2019qqr}, which can enhance the SPS predictions.
DPS may also appreciably contribute in these regions, since separate hard interactions yield less correlated mesons and broaden the invariant-mass and rapidity separation distributions in the factorized description~\cite{CMS:2020qwa}.
These possibilities require quantitative assessment for double $\Upsilon(1S)$ hadroproduction before their relative importance can be established.
Our present NLO CS SPS predictions, including their changes in shape, provide valuable input for future joint analyses of the integrated and differential cross sections aimed at discriminating between SPS and DPS and extracting $\sigma_{\mathrm{eff}}$.

To deepen our understanding of the double $\Upsilon$ hadroproduction mechanism, we also investigated the $p_T^{\Upsilon}$ distribution at NLO in $\alpha_s$ and the $p_T^{\Upsilon\Upsilon}$ distribution, which comes about because the meson pair can recoil against an additional parton, breaking the back-to-back kinematics at LO.
Our results show that the QCD corrections significantly modify the LO lineshapes, with the threshold contributions being suppressed and the tails being enhanced.

\subsection*{Acknowledgments}

This work was supported in part by the German Research Foundation DFG through Grants No.\ KN 365/13-2 and 365/14-2, and by Fundamental Research Funds for the Central Universities through Grant No.\ buctrc202432.


\end{document}